\documentclass[12pt]{iopart}
\usepackage{amssymb}
\begin{document}
\title{Friedmann Equation for Brans Dicke Cosmology}
\author{M. Ar\i k, M. \c{C}al\i k and M. B. Sheftel}
\begin{abstract}

In the context of Brans-Dicke scalar tensor theory of gravitation,
the cosmological Friedmann equation which relates the expansion rate
$H$ of the universe to the various fractions of energy density is
analyzed rigorously. It is shown that Brans-Dicke scalar tensor
theory of gravitation brings a negligible correction to the matter
density component of Friedmann equation. Besides, in addition to
$\Omega _{\Lambda }$ and $\Omega _{M}$ in standard Einstein
cosmology, another density parameter, $\Omega _{_{\Delta }}$, is
expected by the theory inevitably. Some cosmological consequences
of such non-familiar case are examined as far as recent
observational results are concerned. Theory  implies that if $\Omega _{_{\Delta }}$
is found to be nonzero, data can favor this model and hence this theory turns out to be the most powerful
candidate in place of the standard Einstein cosmological model with cosmological constant. Such replacement
will enable more accurate predictions for the rate of change of
Newtonian gravitational constant in the future.
\end{abstract}

\address{Bo\~{g}azi\c{c}i Univ., Dept. of Physics, Bebek, Istanbul, Turkey}
\address{Dogus Univ., Dept. of Sciences, Acibadem, Zeamet Street No: 21
34722 Kadikoy, Istanbul, Turkey}
\eads{\mailto{metin.arik@boun.edu.tr},
\mailto{mcalik@dogus.edu.tr},
\mailto{mikhail.sheftel@boun.edu.tr}}

Recent observational data have strongly confirmed that we live in an
accelerating universe \cite{correl2} and have made it possible to determine
the composition of the universe \cite{bernardis}-\cite{peebles}. According
to these observations, nearly seventy percent of the energy density in the
universe is unclustered (dark energy) and has negative pressure by which it is driving an
accelerated expansion \cite{Efstathiou}-\cite{T. Roy}. Furthermore, the
energy density of the vacuum is much smaller than the estimated values so
far. By itself, acceleration seems to be much more understandable in the
context of general relativity (cosmological constant) \cite{correl} and
quantum field theory (quantum zero point energy); however, the very small
and non-zero energy scale implied by the observations is not quite
comprehensible. Because of these conceptual problems associated with the
cosmological constant \cite{padmn}-\cite{gon}, alternative treatments to the
problem have been produced and they are being used widely in the literature
nowadays \cite{4}-\cite{8}. For more detailed explanation about
a number of approaches proposed so far and recent progress
made towards understanding the nature of this dark energy see \cite{copeland}. In some
of these treatments, mostly, a scalar field $\phi
$ is considered together with a suitably chosen $V({\phi })$ to make the
vacuum energy vary with time.

To get a model in which
the current value of the cosmological constant can be expressed in a more
natural way; namely, without need of any fine tuning, in the literature, there exist
a number of studies on accelerated models in Brans Dicke theory \cite{o.arias}-%
\cite{Das}. For example, Sen \textit{\ et al} \cite{11} have found the
potential relevant to power law expansion in Brans-Dicke (BD) cosmology
whereas Ar\i k and \c{C}al\i k \cite{12} have shown that BD theory of
gravity with the standard mass term potential ($1/2$)$m^{2}\phi ^{2}$ is a
beneficial theory in both explaining the rapid primordial and slow late-time
inflation. In this regard, we have chosen the underlying theory as a scalar
tensor theory, especially, BD scalar tensor theory of gravitation since
scalar-tensor theories are the most serious alternative to standard general
relativity. The theory is parameterized by a dimensionless constant $\omega $%
, as $\omega \rightarrow \infty $ BD theory approaches the Einstein
theory \cite{will}. In BD
model, Lagrangian is defined by as in the following form
\begin{equation}
L_{BD}=\sqrt{-g}\left[ \left( \varphi R-\omega \frac{1}{\varphi }g^{\mu \nu
}\partial _{\mu }\varphi \partial _{\nu }\varphi \right) +L_{M}(\Psi )\right]
,  \label{protypebd}
\end{equation}%
where the dimensionless $\omega $ is the only parameter of the theory, and $%
L_{M}$ is the matter Lagrangian. According to our metric convention, (+ - -
-), Equation (\ref{protypebd}) turns out to be
\begin{equation}
L_{BD}=\sqrt{-g}\left[ \left( -\varphi R+\omega \frac{1}{\varphi }g^{\mu \nu
}\partial _{\mu }\varphi \partial _{\nu }\varphi \right) +L_{M}(\Psi )\right]
.  \label{protypebda}
\end{equation}%
In particular, it is expected that $\varphi $ $(t,$ $\vec{x})$ is spatially
uniform and evolves slowly only with cosmic time $t$ such that $\varphi $ $%
(t,$ $\vec{x})\rightarrow $ $\varphi (t)$. As another point, the second term
on the right hand side of (\ref{protypebd}) appears to be kinetic term of
the scalar field but it is in an unlikely form, since the presence of the $%
\varphi ^{-1}$ which seems to indicate a singularity, and the presence of
the coupling constant in multiplicative form is undesired. However, whole
term can be transformed into the standard canonical form by re-defining the
scalar field $\varphi $ by introducing a new field $\phi $, and a new
constant $\epsilon $ in such a way that%
\begin{equation}
\varphi =\frac{1}{8\omega }\phi ^{2}  \label{nwfild}
\end{equation}%
where $\epsilon =\frac{1}{4\omega }$.

In this new form, BD Lagrangian is redefined as
\begin{equation}
L_{BD}=\sqrt{-g}\left[ -\frac{1}{8\omega }\phi ^{2}R+\frac{1}{2}g^{\mu \nu
}\partial _{\mu }\phi \partial _{\nu }\phi +L_{M}(\Psi )\right] ,
\label{BDLagr}
\end{equation}%
where the signs of the non-minimal coupling term and the kinetic energy term
are properly adopted to $(+---)$ metric signature in such a way that as $%
g^{\mu \upsilon }\sim \eta ^{00}$, the kinetic term, $\frac{1}{2}\,g^{\mu
\upsilon }\,\partial _{\mu }\phi \,\partial _{\nu }\phi $ becomes $\frac{1}{2%
}\,\dot{\phi}^{2}$. Since the definition of $\omega $ is not changed, the
limit $\omega $ approaching infinity is the same in both cases. In this limit $\varphi $ remains
constant whereas, as one can see from the above relation the limit of $\phi
^{2}$ is singular. Since the matter part of lagrangian density $L_{M}$ does
not contain neither $\phi $ nor $\omega $ it does not take any part in this
transformation and therefore is unchanged. Present limits of the constant $\omega $ based on
time-delay experiments \cite{C.M Will}-\cite{reasanberg} require $\omega
>10^{4}\gg 1$. Besides, since these theories have been found as the low
energy limit of string theory and they provide appealing models for
inflation, scalar tensor theories enable an interesting arena where the
standard model can be tested. Hence, in this work, we aim to calculate the
corrections, in the context of BD cosmology, to the famous Friedmann
equation
\begin{equation}
\left( \frac{H}{H_{0}}\right) ^{2}=\Omega _{\Lambda }+\Omega _{R}\left(
\frac{a_{0}}{a}\right) ^{2}+\Omega _{M}\left( \frac{a_{0}}{a}\right) ^{3}\
\label{frd}
\end{equation}%
which relates the expansion rate $H=\dot{a}/a$ of the universe to the energy
density. The fractional density parameter, $\Omega_{i} $, is defined as
the ratio of the energy density to the critical energy density which is a
special required density in order to make the geometry of the universe flat.
The Friedmann equation is used for fitting the Hubble parameter, $H$, to the
measured density parameters ($\Omega _{\Lambda }$, $\Omega _{R}$, $\Omega
_{M}$) of the universe in such a way that $\Omega _{\Lambda }+$ $\Omega
_{R}+ $ $\Omega _{M}=1$. According to recent observational results for the
present universe, we have $\Omega _{\Lambda }\simeq 0.75$, $\Omega
_{M}\simeq 0.25$, $\Omega _{R}\simeq 0$ \cite{knop}. In the light of these
values, one can conclude that the universe is mostly filled with
non-baryonic matter and it seems that this non baryonic matter is
responsible for the expansion of the universe solely. In the context of (BD)
theory \cite{13} whose scalar potential term consists of only a mass term
 and matter fields, the
action in the canonical form is given by%
\begin{equation}
S=\int d^{4}x\,\sqrt{g}\,\left[ -\frac{1}{8\omega }\,\phi ^{2}\,R+\frac{1}{2}%
\,g^{\mu \upsilon }\,\partial _{\mu }\phi \,\partial _{\nu }\phi -\frac{1}{2}%
m^{2}\phi ^{2}+L_{M}\right] .  \label{action*}
\end{equation}%
The signs of non-minimal coupling term and the kinetic energy term are
properly adopted to $(+---)$ metric signature. As in the cosmological
approximation, $\phi $ is spatially uniform, but varies slowly with time. As
long as the dynamical scalar field $\phi $ varies slowly, $G_{eff}$, the
effective gravitational constant, is defined as $G_{eff}^{-1}=\frac{2\pi }{%
\omega }\phi ^{2}$ by replacing the non-minimal coupling term $\frac{1}{%
8\omega }\phi ^{2}\,R$ with the Einstein-Hilbert term $\frac{1}{16\Pi G_{N}}%
R $ where $R$ is the Ricci scalar in Einstein relativity. In natural
units where $c=\hbar =1$, we define the Planck-length, $L_{P}^{2}=\omega
/2\pi \phi _{0}^{2}$ where $\phi _{0}$ is the present value of the scalar
field $\phi $. Thus, the dimension of the scalar field is $L^{-1}$ so that
the dimension of $G_{eff}$ is $L^{2}$. $L_{M}$ is decoupled from $\phi $ as
was assumed in the original BD theory. Hence, considering $\phi $ does not
couple to $L_{M}$ as a matter field, we may consider a classical perfect
fluid with the energy-momentum tensor $T_{\nu }^{\mu }=diag\left( \rho
,-p,-p,-p\right) $ where $p$ is the pressure. The gravitational field
equations derived from the variation of the action (\ref{action*}) with
respect to the Robertson-Walker metric are
\begin{equation}
\frac{3}{4\omega }\,\phi ^{2}\,\left( \frac{\dot{a}^{2}}{a^{2}}+\frac{k}{%
a^{2}}\right) -\frac{1}{2}\,\dot{\phi}^{2}-\frac{1}{2}\,m^{2}\,\phi ^{2}+%
\frac{3}{2\omega }\,\frac{\dot{a}}{a}\,\dot{\phi}\,\phi =\rho _{M}
\label{des}
\end{equation}%
\begin{equation}
\frac{-1}{4\omega }\phi ^{2}\left( 2\frac{\ddot{a}}{a}+\frac{\dot{a}^{2}}{%
a^{2}}+\frac{k}{a^{2}}\right) -\frac{1}{\omega }\,\frac{\dot{a}}{a}\,\dot{%
\phi}\,\phi -\frac{1}{2\omega }\,\ddot{\phi}\,\phi -\left( \frac{1}{2}+\frac{%
1}{2\omega }\right) \,\dot{\phi}^{2}+\frac{1}{2}\,m^{2}\,\phi ^{2}=p_{M}
\label{pres}
\end{equation}%
\begin{equation}
\ddot{\phi}+3\,\frac{\dot{a}}{a}\,\dot{\phi}+\left[ m^{2}-\frac{3}{2\omega }%
\left( \frac{\ddot{a}}{a}+\frac{\dot{a}^{2}}{a^{2}}+\frac{k}{a^{2}}\right) %
\right] \,\phi =0  \label{fi}
\end{equation}%
where $k$\ is the curvature parameter with $k=-1$, $0$, $1$\ corresponding
to open, flat, closed universes respectively and $a\left( t\right) $ is the
scale factor of the universe (dot denotes $d/dt$), $M$ denotes everything
except the $\phi $ field. The right hand side of the $\phi $ equation (\ref%
{fi}) is set to be zero due to the assumption that the matter Lagrangian $%
L_{M}$ is independent of the scalar field $\phi $.  Since the matter lagrangian affects
the first two equations (\ref{des}-\ref{pres}) and when $R$ is solved from the first two equations
and substituted into the $\phi $ equation (\ref{fi}) this equation depends on the trace of
the energy momentum tensor $T_{\mu \nu }$. Indeed this is how the BD field
equations are sometimes written:
\begin{equation}
\ddot{\phi}\phi +\,\dot{\phi}^{2}-\left( \frac{2\omega }{2\omega +3}\right)
m^{2}\,\phi ^{2}+3\frac{\dot{a}}{a}\dot{\phi}\phi +\left( \frac{2\omega }{%
2\omega +3}\right) \left( 3p-\rho \right) =0.
\end{equation}
But in any case the system of equations and the solutions to the system of
equations are independent of this phenomenon. Instead of working with
the field equations (\ref{des}-\ref{fi}) stated in terms of $\phi (t)$, $%
a(t) $ and their derivatives with respect to the cosmological time $t$. We
define the fractional rate of change of $\phi $ as $F\left( a\right) =\dot{%
\phi}/\phi $ and the Hubble parameter as $H\left( a\right) =\dot{a}/a$, and
rewrite the left hand-side of the field equations (\ref{des}-\ref{fi}) in
terms of $H(a)$, $F(a)$ and their derivatives with respect to the scale size
of the universe $a$ (prime denotes $\frac{d}{da}$)%
\begin{equation}
H^{2}-\frac{2\omega }{3}\,F^{2}+2\,H\,F+\frac{k}{a^{2}}-\frac{2\omega }{3}%
\,m^{2}=\left( \frac{4\omega }{3}\right) \frac{\rho _{M}}{\phi ^{2}}
\label{HF1**}
\end{equation}%
\begin{equation}
H^{2}+\left( \frac{2\omega }{3}+\frac{4}{3}\right) \,F^{2}+\frac{4}{3}\,H\,F+%
\frac{2a}{3}\,\left( H\,H^{\prime }+H\,F^{\prime }\right) +\frac{k}{3a^{2}}-%
\frac{2\omega }{3}\,m^{2}=\left( \frac{-4\omega }{3}\right) \frac{p_{M}}{%
\phi ^{2}}  \label{HF2****}
\end{equation}%
\begin{equation}
H^{2}-\frac{\omega }{3}\,F^{2}-\omega \,H\,F+a\left( \frac{H\,H^{\prime }}{2}%
-\frac{\omega }{3}\,HF^{\prime }\right) +\frac{k}{2a^{2}}-\frac{\omega }{3}%
\,m^{2}=0.  \label{hf3}
\end{equation}%
From these three equations it can be shown that the continuity equation for
the matter-energy excluding the BD scalar field is also satisfied with the
help of the $\phi $ equation (\ref{hf3})%
\begin{equation}
\dot{\rho _{M}}+3\left( \frac{\dot{a}}{a}\right) (p_{M}+\rho _{M})=0
\label{cont}
\end{equation}%
and hence, instead of considering the $p$ equation (\ref{HF2****}) solely as
one of the dynamical equations to be satisfied, we choose continuity
equation in addition to the density equation and the $\phi $ equation to be
satisfied in any cosmological case we want to explain. That is because once
the continuity equation is satisfied than $p$ equation must already be
satisfied automatically provided that $\dot{a}$ is nonzero. To eliminate the
$\phi $ dependence in (\ref{HF1**}), we take the time derivative of both
sides of the $\rho $ equation and after some rearrangements, we get (\ref%
{HF1**}) purely in terms of $H(a)$, $F(a)$, $\rho (a)$ and their derivatives
with respect to $a$.
\begin{eqnarray}
H^{\prime }(H^{2}+HF)+F^{\prime }(H^{2}-\frac{2\omega }{3}HF)  \nonumber \\
\lo=\frac{H^{3}}{2}\left( \frac{\rho ^{^{\prime }}}{\rho }\right) +\frac{%
2\omega }{3a}F^{3}+H^{2}F\left[ \left( \frac{\rho ^{^{\prime }}}{\rho }%
\right) -\frac{1}{a}\right] +F^{2}H\left[ -\frac{2}{a}-\frac{\omega }{3}%
\left( \frac{\rho ^{^{\prime }}}{\rho }\right) \right]  \nonumber \\
+\frac{k}{a^{2}}\left[ H\left( \left( \frac{\rho ^{^{\prime }}}{2\rho }%
\right) +\frac{1}{a}\right) -\frac{F}{a}\right] -\omega m^{2}\left[ H\left(
\frac{\rho ^{^{\prime }}}{3\rho }\right) -\frac{2F}{3a}\right] .
\label{rho1}
\end{eqnarray}%
After rewriting (\ref{hf3}) in the following form%
\begin{equation}
3aHH^{^{\prime }}-2\omega HaF^{^{\prime }}=-6H^{2}+2\omega F^{2}+6\omega HF-%
\frac{3k}{a^{2}}+2\omega m^{2}  \label{newfi}
\end{equation}
we solve (\ref{rho1}, \ref{newfi}) for $H^{\prime }$, $F^{\prime }$ and get
the general form of the solution in the sense that once the curvature
constant $k$ and energy density in terms of $a$ is given than $H$ and $F$
can be solved from the following equations:
\begin{eqnarray}
\lo H^{\prime }=\frac{\left[ \omega a(\rho ^{\prime }/\rho )-6\right] }{%
\left( 2\omega +3\right) aH}H^{2}-\frac{\left[ 4\omega ^{2}+2\omega
+2a\omega ^{2}(\rho ^{\prime }/3\rho \right] }{\left( 2\omega +3\right) aH}%
F^{2}+\frac{\left[ 8\omega +2a\omega (\rho ^{\prime }/\rho )\right] }{\left(
2\omega +3\right) aH}HF  \nonumber \\
\lo -\frac{\left[ 2\omega ^{2}a(\rho ^{\prime }/3\rho )-2\omega \right] }{%
\left( 2\omega +3\right) aH}m^{2}+k\frac{\left[ 2\omega +\omega a(\rho
^{\prime }/2)-3\right] }{\left( 2\omega +3\right) a^{3}H}  \label{H' newform}
\end{eqnarray}%
\begin{eqnarray}
F^{\prime }=\frac{\left[ 3a(\rho ^{\prime }/2\rho )+6\right] }{\left(
2\omega +3\right) aH}H^{2}-\frac{\left[ 8\omega +a\omega (\rho ^{\prime
}/\rho )+6\right] }{\left( 2\omega +3\right) aH}F^{2}-\frac{\left[ 6\omega
-3a(\rho ^{\prime }/\rho )-3\right] }{\left( 2\omega +3\right) aH}HF
\nonumber \\
\lo-\frac{\left[ \omega a(\rho ^{\prime }/\rho )+2\omega \right] }{\left(
2\omega +3\right) aH}m^{2}+k\frac{\left[ 6+3a(\rho ^{\prime }/2\rho )\right]
}{\left( 2\omega +3\right) a^{3}H}.  \label{F'newform}
\end{eqnarray}%
Hence, in the present epoch, to discover how the Hubble parameter $H$
changes with the scale size of the universe $a$, we assume that the present
universe is mostly flat and it necessarily obeys the $p_{M}=0$ equation of
state . Using (\ref{cont}), we find that the energy density $\rho $ evolves
with $a$ in the same manner as in standard Einstein cosmology when the
universe is solely governed by matter,
\begin{equation}
\rho =\frac{C}{a^{3}}  \label{rcoz}
\end{equation}%
where $C$ is an integration constant. Setting $k=0$ and inserting this
energy density into (\ref{H' newform}, \ref{F'newform}), we get the
following form of the equations to be solved:
\begin{equation}
H^{^{\prime }}=\frac{-1}{H(2\omega +3)a}\left[ 3(2+\omega )H^{2}+2\omega
(\omega +1)F^{2}-2\omega HF-2\omega (\omega +1)m^{2}\right]  \label{enewH'}
\end{equation}%
\begin{equation}
F^{^{\prime }}=\frac{1}{H(2\omega +3)a}\left[ \frac{3}{2}H^{2}-\left(
5\omega +6\right) F^{2}-6(1+\omega )HF+\omega m^{2}\right] .  \label{enewF'}
\end{equation}
With the transformation $u=\left( \frac{a_{0}}{a}\right) ^{\alpha }$ , we
rewrite (\ref{enewH'}, \ref{enewF'}) in terms of $H(u)$, $F(u)$ and their
derivatives with respect to $u$%
\begin{equation}
\fl\frac{dH}{du}=\frac{1}{\alpha H(2\omega +3)u}\left[ 3\left( 2+\omega
\right) H^{2}+2\omega \left( \omega +1\right) F^{2}-2\omega HF-2\omega
(\omega +1)m^{2}\right]  \label{H' (u)}
\end{equation}

\begin{equation}
\fl\frac{dF}{du}=\frac{-1}{\alpha H(2\omega +3)u}\left[ \frac{3}{2}%
H^{2}-\left( 5\omega +6\right) F^{2}-6(1+\omega )HF+\omega m^{2}\right] .
\label{F'(u)}
\end{equation}%
Since these coupled equations are hard enough to be solved analytically for $%
H$ and $F$, our approach is to determine a perturbative solution in which
both $H$ and $F$ vary about some constants $H_{\infty }$ and $F_{\infty }$
respectively:
\begin{equation}
H=H_{\infty }+H_{1}u+H_{2}u^{2}+...  \label{proH}
\end{equation}%
\begin{equation}
F=F_{\infty }+F_{1}u+F_{2}u^{2}+...  \label{proF}
\end{equation}%
where $H_{\infty }$, $F_{\infty }$, $H_{1}$, $F_{1}$, $\alpha $, are all
constants to be determined from the theory and from fitting the Hubble
parameter, $H$, to the measured density parameters ($\Omega _{\Lambda }$, $%
\Omega _{R}$, $\Omega _{M}$) of the universe via Friedmann equation.
Plugging this perturbative solution into (\ref{H' (u)}, \ref{F'(u)}) and
keeping only the zeroth, first, second order terms of $u$ and neglecting
higher terms, we end up with two sets of solutions in the zeroth order%
\begin{equation}
H_{\infty }=\frac{\sqrt{\omega }\left( 2\omega +2\right) m}{\sqrt{\left(
6\omega ^{2}+17\omega +12\right) }}\,;\;F_{\infty }=\frac{H_{\infty }}{%
2(\omega +1)}  \label{solHandF}
\end{equation}%
and
\begin{equation}
H_{\infty }=\frac{2\sqrt{3\omega }\,m}{3\sqrt{3\omega +4}}\,;\;F_{\infty }=%
\frac{3}{2}H_{\infty }.  \label{so2HandF}
\end{equation}%
Comparing the first order terms of $u$, on the other hand, provides two
linearly dependent equations for which the only possible solution is the
trivial solution of $H_{1}=0$ and $F_{1}=0$,
\begin{equation}
\fl\left\{ \left[ 6(\omega +2)-\alpha (2\omega +3)\right] H_{\infty
}-2\omega F_{\infty }\right\} H_{1}+\left[ -2\omega H_{\infty }+4\omega
(\omega +1)F_{\infty }\right] F_{1}=0  \label{h1f1-1}
\end{equation}%
\begin{equation}
\fl\left[ -3H_{\infty }+6(\omega +1)F_{\infty }\right] H_{1}+\left\{ \left[
6\left( \omega +1\right) -\alpha (2\omega +3)\right] H_{\infty }+2(5\omega
+6)F_{\infty }\right\} F_{1}=0.  \label{h1f1-2}
\end{equation}%
Since the solution in which $H_{1}$ and $F_{1}$ are nonzero is much more
plausible for our aim, the coefficient matrix is properly constructed from (%
\ref{h1f1-1}, \ref{h1f1-2}) and its determinant is set to be zero to get the
value of $\alpha $ for which $H_{1}$ and $F_{1}$ need not be zero
simultaneously. We get two different $\alpha $ values
\begin{equation}
\alpha =3+\frac{1}{1+\omega }  \label{good}
\end{equation}%
and
\begin{equation}
\alpha \sim \sqrt{\omega }  \label{bad}
\end{equation}%
corresponding to the solution sets (\ref{solHandF}) and (\ref{so2HandF})
respectively. In this regard, we note two things here:

\begin{itemize}
\item Concerning the solution of $H$ we seek for, the solution (\ref{good})
is much more precious than the solution (\ref{bad}) which approaches to
infinity as $\omega $ becomes infinitely large. On the other hand, in the
same limit, (\ref{good}) gives $\alpha =3$ which is the well known term in a
matter dominated universe solution of standard Einstein cosmology.

\item The correction factor $1/(1+\omega )$ in the solution (\ref{good}) is
solely coming from the exact solutions of the field equations of BD theory.
\end{itemize}

Two linearly dependent equations are available when one compares the second
order terms of $u$;
\begin{eqnarray}
&&\left\{ \left[ 6(\omega +2)-2\alpha (2\omega +3)\right] H_{\infty
}-2\omega F_{\infty }\right\} H_{2}+\left[ 4\omega (\omega +1)F_{\infty
}-2\omega H_{\infty }\right] F_{2}  \nonumber \\
&=&\left[ \alpha (2\omega +3)-3(\omega +2)\right] H_{1}^{2}-2\omega (\omega
+1)F_{1}^{2}+2\omega H_{1}F_{1}  \label{H2F2a}
\end{eqnarray}%
\begin{eqnarray}
&&\left[ 3H_{\infty }-6(\omega +1)F_{\infty }\right] H_{2}+\left\{ \left[
2\alpha (2\omega +3)-6\left( \omega +1\right) \right] H_{\infty }-2(5\omega
+6)F_{\infty }\right\} F_{2}  \nonumber \\
&=&-\frac{3}{2}H_{1}^{2}+F_{1}^{2}+\left[ 6\left( \omega +1\right) -\alpha
(2\omega +3)\right] H_{1}F_{1}.  \label{H2F2b}
\end{eqnarray}%
Letting $\alpha =3+1/(\omega +1)$ and $F_{\infty }=H_{\infty }/2(\omega +1)$
in (\ref{H2F2a}, \ref{H2F2b}) gives $H_{2}$ and $F_{2}$ only in terms of $%
H_{\infty }$, $H_{1}$, $F_{1}$;
\begin{equation}
H_{2}=\frac{\left[ -(3\omega ^{2}+8\omega +6)H_{1}^{2}+2\omega (\omega
+1)^{2}F_{1}^{2}-2\omega (\omega +1)H_{1}F_{1}\right] }{(3\omega +4)(2\omega
+3)H_{\infty }}  \label{h2exact}
\end{equation}%
\begin{equation}
F_{2}=\frac{\left[ \frac{-3(\omega +1)}{2}H_{1}^{2}+(\omega
+1)F_{1}^{2}-(5\omega +6)H_{1}F_{1}\right] }{(3\omega +4)(2\omega
+3)H_{\infty }}.  \label{F2}
\end{equation}%
Hence, with these perturbation constants found from theory, we can express $H
$ and $F$ as
\begin{equation}
H=H_{\infty }+H_{1}\left( \frac{a_{0}}{a}\right) ^{3+\frac{1}{\omega +1}%
}+H_{2}\left( \frac{a_{0}}{a}\right) ^{6+\frac{2}{\omega +1}}+...
\label{Hrst}
\end{equation}%
\begin{equation}
F=F_{\infty }+F_{1}\left( \frac{a_{0}}{a}\right) ^{3+\frac{1}{\omega +1}%
}+F_{2}\left( \frac{a_{0}}{a}\right) ^{6+\frac{2}{\omega +1}}+...
\label{Frst}
\end{equation}%
where (\ref{solHandF}) gives
\begin{equation}
H_{\infty }=\left[ 2\left( \omega +1\right) \sqrt{\omega }m\right] /\sqrt{%
\left( 6\omega ^{2}+17\omega +12\right) }  \label{Hinf}
\end{equation}%
and
\begin{equation}
F_{\infty }=\left( \sqrt{\omega }m\right) /\sqrt{\left( 6\omega
^{2}+17\omega +12\right) }.  \label{Finf}
\end{equation}%
To proceed one step further, we rewrite the standard Friedmann equation (\ref%
{frd}) with the extra term having new density parameter, $\Omega _{\Delta }$%
,
\begin{equation}
\left( \frac{H}{H_{0}}\right) ^{2}=\Omega _{\Lambda }+\Omega _{M}\,\left(
\frac{a_{0}}{a}\right) ^{3+\frac{1}{\omega +1}}+\Omega _{\Delta }\,\left(
\frac{a_{0}}{a}\right) ^{6+\frac{2}{\omega +1}}  \label{frd in Brans-dicke}
\end{equation}%
and we fit all theory parameters to the observational density parameters;
\begin{equation}
\Omega _{\Lambda }=\frac{H_{\infty }^{2}}{H_{\Sigma }^{2}},
\label{omegalamda}
\end{equation}%
\begin{equation}
\Omega _{M}=\frac{2H_{\infty }H_{1}}{H_{\Sigma }^{2}},  \label{omegamatter}
\end{equation}%
\begin{equation}
\Omega _{\Delta }=\frac{H_{1}^{2}+2H_{\infty }H_{2}}{H_{\Sigma }^{2}},
\label{omegadelta}
\end{equation}%
where%
\begin{equation}
H_{\Sigma }^{2}=H_{\infty }^{2}+2H_{\infty }(H_{1}+H_{2})+H_{1}^{2}.
\label{omegasigma}
\end{equation}%
With these relations above and the constraint $\Omega _{\Lambda }$ $+$ $%
\Omega _{M}+\Omega _{_{\Delta }}=1$, we can express theoretical parameters $%
H_{1}$, $H_{2}$ in terms of the observational density parameters $\Omega
_{\Lambda }$, $\Omega _{M}$ and $H_{\infty }$:%
\begin{equation}
H_{1}=\frac{\Omega _{M}}{2\sqrt{\Omega _{\Lambda }}}H_{\infty }  \label{he*}
\end{equation}%
\begin{equation}
H_{2}=\left[ 2-\left( \sqrt{2\Omega _{\Lambda }}+\frac{\Omega _{M}}{\sqrt{%
2\Omega _{\Lambda }}}\right) ^{2}\right] H_{\infty }.  \label{he2}
\end{equation}%
Now at this stage, investigation of two cases for $\Omega _{\Delta }$ can be
meaningful:

We first set $\Omega _{\Delta }\simeq 0$, ie; $\Omega _{M}+\Omega _{\Lambda
}=1$ consistent with today's universe density compositions and $H_{2}\neq 0$%
. By using recent observational results on density parameters $\Omega
_{M}\simeq 0.27$, $\Omega _{\Lambda }\simeq 0.73$ \cite{WMAP}-\cite{WMAP2} together
with (\ref{he*}) and (\ref{he2}), we determine
\begin{equation}
H_{1}=\frac{0.23}{2\sqrt{0.73}}H_{\infty }\simeq 0.15H_{\infty },
\label{H1 delta=0}
\end{equation}%
\begin{equation}
H_{2}=-\frac{H_{1}^{2}}{2H_{\infty }}\simeq -\frac{\left( 0.15H_{\infty
}\right) ^{2}}{2H_{\infty }}\simeq -0.01H_{\infty }<0.  \label{H2 delta=0}
\end{equation}%
Now, if (\ref{Hrst}) is satisfied for $H=H_{0}$, we get%
\begin{equation}
H_{\infty }\simeq 0.88H_{0}  \label{Hinf delta=0}
\end{equation}%
where $H_{0}$ is the present value of the Hubble parameter \cite{Freedmann}
and we may estimate mass of BD scalar field, $m,$ for a fixed $H_{0}$ as $%
m\lessapprox 10^{-2}H_{0}$ by using the relation $H_{\infty }\simeq
0.88H_{0}\simeq 0.82m\,\omega ^{1/2}$ for $\omega \rightarrow \infty $.
Furthermore, by using (\ref{h2exact}), (\ref{F2}), (\ref{H1 delta=0}), (\ref%
{H2 delta=0}) simultaneously, we get $F_{1}\simeq 0.08H_{\infty }\,/\omega $%
, $F_{2}\simeq -0.04H_{\infty }\,/\omega $ for $\omega \rightarrow \infty $.
Remembering that $F_{\infty }\simeq H_{\infty }\,/2\omega $, we may say that
$F_{\infty }$ in (\ref{Frst}) is the dominating term in today's universe.
This shows us that similar to the expansion rate of the universe $H$, the
rate of change of the Newtonian gravitational constant has approached the
asymptotic regime.

On the contrary, when we set $\Omega _{\Delta }\neq 0$ but $\Omega _{\Delta
}>$ $0$ ie; $\Omega _{M}+\Omega _{\Lambda }+\Omega _{\Delta }=1$ together
with $H_{2}\simeq 0$, from (\ref{he2}), we immediately get the relation%
\begin{equation}
\Omega _{M}=2\sqrt{\Omega _{\Lambda }}\left( 1-\sqrt{\Omega _{\Lambda }}%
\right) ,  \label{DENS}
\end{equation}%
and via this relation, we are able to estimate $\Omega _{\Lambda }$ and $%
\Omega _{_{_{\Delta }}}$as in the following portions if the recent measured
observational result on matter density parameter $\Omega _{M}\simeq 0.27$ \cite{WMAP} is
kept fixed,
\begin{eqnarray}
\Omega _{\Lambda } &\simeq &0.71  \nonumber \\
\Omega _{_{_{\Delta }}} &\simeq &0.02.  \label{prdic}
\end{eqnarray}
To figure out the reliance of such theory based density parameters
 and to show how this model can pass the
observational constraint of the SNIa data \cite{SNIa}, we have compared our model of $%
\Omega _{M}=2\sqrt{\Omega _{\Lambda }}\left( 1-\sqrt{\Omega _{\Lambda }}%
\right) $ proposing $\Omega _{_{_{\Delta }}}\simeq 0.02$, $\Omega _{M}\simeq
0.27$ and $\Omega _{\Lambda }\simeq 0.71$ with LCDM model of $\ \Omega
_{\Lambda }=1-$ $\Omega _{M}$ proposing $\Omega _{M}=0.276\pm 0.026$ and $%
\Omega _{\Lambda }$ $=0.724\pm 0.026$ according to wmap+SNIa data (\textbf{%
ref}). From such comparison, it can be seen that in an asymptotic regime,
although $\Omega _{\Delta }\neq 0$, SNIa data do not exclude our
model as far as the range of uncertainties proposed by WMAP data team are valid.
Besides, under this non-familiar
case, we get other theory parameters $F_{1}\simeq 0.2H_{\infty }\,/\sqrt{%
\omega }$, $F_{2}\simeq -7\times 10^{-3}H_{\infty }\,/\omega $ as $\omega
\rightarrow \infty $ . Remembering that $F_{\infty }\simeq H_{\infty
}\,/2\omega $, we see that $F_{1}$ is the dominating term in (\ref{Frst}),
namely, the rate of change of the Newtonian gravitational constant $\left(
\dot{G}_{N}/G_{N}\right) $ has not approached to the asymptotic regime yet.
However, theory predicts that when the size of the universe exceeds $a\gg
0.6\omega ^{1/6}a_{0}$, then the term $F_{\infty }$ will become dominant so
that asymptotic regime will be satisfied for $\left( \dot{G}%
_{N}/G_{N}\right) $.

As a last case, we investigate the possibility that $\Omega _{_{\Delta }}<0$
theoretically, ie; $\Omega _{\Lambda }$ $+$ $\Omega _{M}>1$. According to
SNIa data, this case is important in the sense that if theory assures that $%
\Omega _{_{\Delta }}<0$ then this will imply that data favors this model
instead of standard Einstein cosmological model with cosmological constant.
However, when we use (\ref{h2exact})and (\ref{omegadelta}) simultaneously
 to make $\Omega _{_{\Delta }}<0$ we
see that for $F_{1}=H_{1}/2(\omega +1)$, $H_{2}$ attains its most negative
value of $H_{2}=-H_{1}^{2}/2H_{\infty }$ and for this value of $H_{2}$, $\Omega _{_{\Delta }}$ exactly
is being equal to zero instead of being equal to negative value. With this result we note that BD theory
can not be forced  to have $\Omega _{_{\Delta }}<0$ so that data can favor
this model instead of standard Einstein cosmological model with cosmological constant.

Hence, in the light of these examinations on $\Omega _{_{\Delta }}$, we may
conclude that measurement of $\Omega _{_{\Delta }}$ will be important in two
respects. Firstly, if $\Omega _{_{\Delta }}$ is found to have positive
value, this will indicate that data can favor this model instead of standard
Einstein cosmological model with cosmological constant. Secondly, if that is so,
making much more accurate predictions for the rate of change of $G_{N}$
could be plausible.

We would like to thank Burak Kaynak for his computational aid which provided
us to be faster in seing the results of the work and Prof. Engin Arik for
her valuable suggestions and contributions to the paper. This work is
supported by Bogazici University Research Fund, Project no:06B301.

\end{document}